# Increased hole mobility in anti-$ThCr_2Si_2$-type $La_2O_2Bi$ co-sintered with alkaline earth metal oxides for oxygen intercalation and hole carrier doping

Kota Matsumoto,[a] Hideyuki Kawasoko,[a] Noriaki Kimura[b] and Tomoteru Fukumura*[a,c]

[a)] Department of Chemistry, Graduate School of Science, Tohoku University, Sendai 980-8578, Japan.
[b)] Department of Physics, Graduate School of Science; Center for Low Temperature Science, Tohoku University, Sendai 980-8578, Japan.
[c)] Advanced Institute for Materials Research and Core Research Cluster, Tohoku University, Sendai 980-8577, Japan.

* tomoteru.fukumura.e4@tohoku.ac.jp

Metallic anti-$ThCr_2Si_2$-type $RE_2O_2Bi$ (*RE* = rare earth) with Bi square nets show superconductivity while insulating $La_2O_2Bi$ shows high hole mobility, by expanding the *c*-axis length through oxygen intercalation. In this study, alkaline earth metal oxides (CaO, SrO, and BaO) were co-sintered with $La_2O_2Bi$. CaO and BaO served as oxygen intercalants without incorporation of Ca and Ba in $La_2O_2Bi$. On the other hand, SrO served as not only oxygen intercalant but also hole dopant via Sr substitution with La in $La_2O_2Bi$. The oxygen intercalation and hole doping resulted in expansion of the *c*-axis length, contributing to improved electrical conduction. In addition, the hole mobility was enhanced up to 150 $cm^2V^{-1}s^{-1}$ in $La_2O_2Bi$, which almost doubles the mobility in previous study.

## Introduction

Square net compounds attract a long-standing interest in broad fields of material science.[1–4] Recently, Bi square net compounds have been extensively studied due to their fascinating electronic properties such as Dirac/Weyl fermion in $AMnBi_2$ (*A* = alkaline earth or rare earth) and superconductivity in antiferromagnetic $CeNi_{0.8}Bi_2$.[5–8] Bi square net compounds are also expected to be topological materials owing to the unusual negative valence state for Bi.[9] Anti-$ThCr_2Si_2$ type $RE_2O_2Bi$ (*RE* = rare earth) consists of conducting $Bi^{2-}$ square nets and insulating $RE_2O_2$ layers. The Fermi surface contributed by only Bi 6p orbital represents a principal role of the monoatomic Bi square nets in the electronic properties.[10,11] Intriguingly, the electronic properties are significantly influenced by their lattice constants: metal-insulator transition induced by expanding *a*-axis length via *RE* substitution[10] and superconducting transition of metallic $RE_2O_2Bi$ (*RE* = Y, Tb, Dy, Er, and Lu) induced by expanding *c*-axis length via excess oxygen intercalation.[12–14] Recently, insulating $La_2O_2Bi$ with the longest *a*-axis length was transformed into a metallic one with high hole mobility of 85 $cm^2V^{-1}s^{-1}$ by expanding the *c*-axis length via oxygen intercalation.[15]

In this study, $La_2O_2Bi$ was co-sintered with alkaline earth metal oxides (CaO, SrO, and BaO). CaO and BaO served as oxygen intercalants, while SrO served as not only oxygen intercalant but also hole dopant via Sr substitution with La. For all the $La_2O_2Bi$, the *c*-axis length was expanded via oxygen intercalation and Sr substitution, contributing to improved electrical conduction. In addition, the hole mobility increased up to 150 $cm^2V^{-1}s^{-1}$ in spite of the polycrystalline form.

## Experimental

$La_2O_{2+\delta}Bi$ and $(La,Sr)_2O_{2+\delta}Bi$ polycrystals were synthesized by solid-state reaction. $La_2O_3$ (99.99%) and $CaCO_3$ (99.99%) powders were heated at 1273–1373 K in air for 10 hours to remove moisture and to decompose $CaCO_3$ into CaO. La (99.9%), $La_2O_3$, Bi (99.9%), CaO, SrO (98%), and BaO (99%) powders were mixed and pelletized under 10 MPa in a nitrogen-filled glovebox to form nominal compositions of $(La_{1-x}AE_x)_2O_{1.6}Bi$ ($0 \le x_{AE} \le 0.20$) (*AE* = Ca, Sr, Ba), which transformed into stoichiometric or oxygen intercalated $(La_{1-x}AE_x)_2O_2Bi$ after sintering as reported previously.[15] The pellets covered with Ta foils were sintered in evacuated quartz tubes at 773 K for 7.5 hours and then at 1273 K for 10 hours. The sintered products were ground and pelletized under 20 MPa again in the glovebox, followed by the sintering in evacuated quartz tubes at 1273 K for 10 hours. The crystal structures were evaluated at room temperature by X-ray powder diffraction (XRD) using Cu Kα radiation (D8 DISCOVER, Bruker AXS). Rietveld analysis was performed by RIETAN-FP (Ref.16) to identify the crystal phases and their lattice parameters. The crystal structures were drawn with the VESTA.[17] The chemical compositions were evaluated by scanning electron microscope equipped with energy dispersive X-ray spectroscopy (SEM-EDX; S-4300, Hitachi). The transport properties were evaluated by physical property measurement system (PPMS, Quantum Design) and a cryostat equipped with dilution refrigerator (Kelvinox TLM, Oxford).

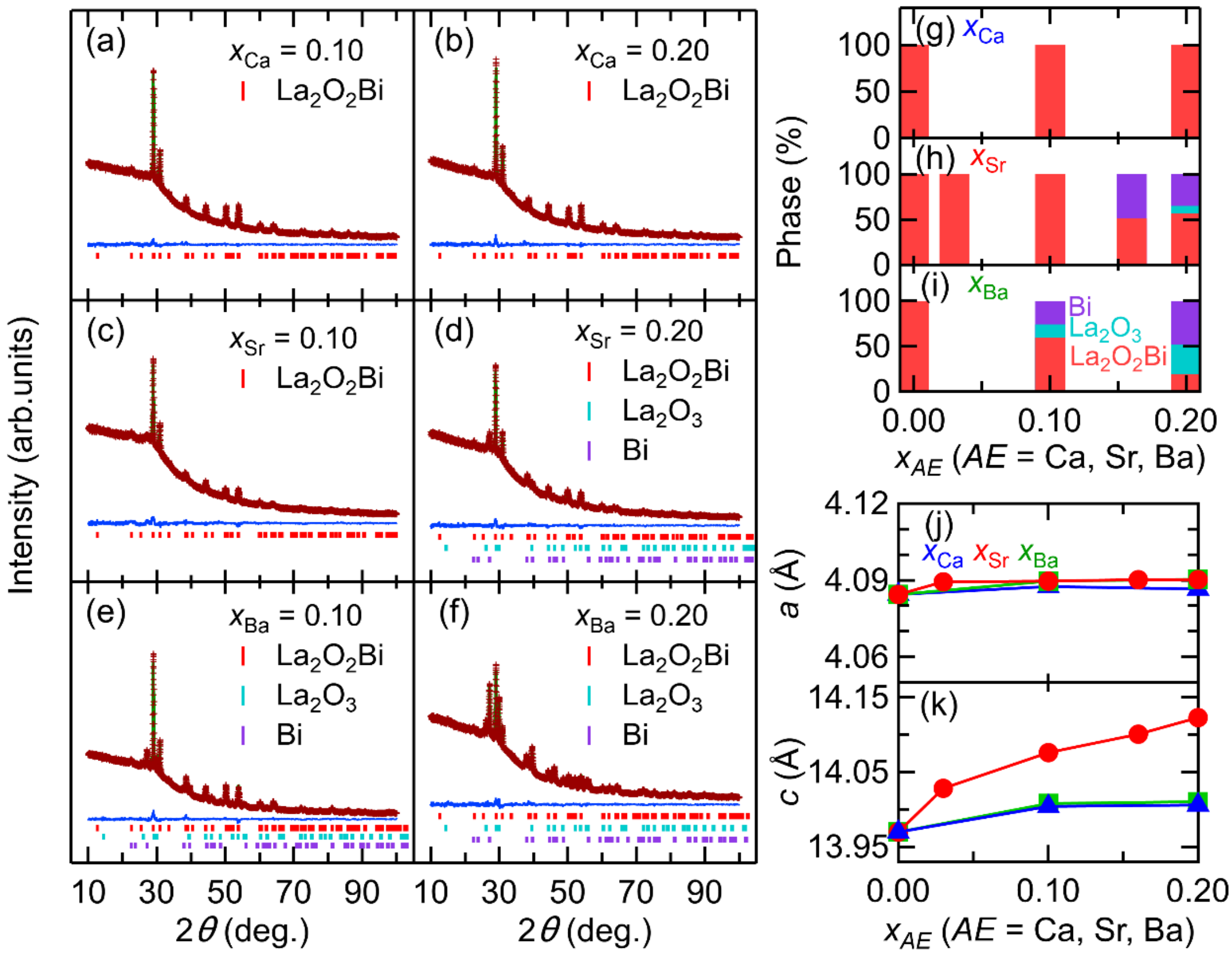

**Fig. 1** XRD patterns and fitting results of Rietveld refinement for (a)–(f) $La_2O_2Bi$ with $x_{Ca}$ = 0.10, 0.20, $x_{Sr}$ = 0.10, 0.20, and $x_{Ba}$ = 0.10, 0.20, respectively. Brown, green, and blue curves denote the measurement data, simulation pattern, and their difference, respectively. (g)–(i) Molar fractions of constituent phases as a function of $x_{AE}$ ($AE$ = Ca, Sr, and Ba), respectively. (j) $a$- and (k) $c$-axis lengths for $La_2O_2Bi$ as a function of $x_{AE}$ ($AE$ = Ca, Sr, and Ba).

**Table 1** Crystal structural parameters for $La_2O_2Bi$ with $x_{Ca}$, $x_{Sr}$, and $x_{Ba}$ obtained by Rietveld analyses.[a]

| | $x_{Ca}$ = 0.10 | $x_{Ca}$ = 0.20 | $x_{Sr}$ = 0.03 | $x_{Sr}$ = 0.10 | $x_{Sr}$ = 0.16 | $x_{Sr}$ = 0.20 | $x_{Ba}$ = 0.10 | $x_{Ba}$ = 0.20 |
|---|---|---|---|---|---|---|---|---|
| Phase | $La_2O_2Bi$ | $La_2O_2Bi$ | $La_2O_2Bi$ | $La_2O_2Bi$ | $La_2O_2Bi$ | $La_2O_2Bi$ | $La_2O_2Bi$ | $La_2O_2Bi$ |
| Space group | *I*4/*mmm* | *I*4/*mmm* | *I*4/*mmm* | *I*4/*mmm* | *I*4/*mmm* | *I*4/*mmm* | *I*4/*mmm* | *I*4/*mmm* |
| $a$ (Å) | 4.0876(1) | 4.0866(1) | 4.0894(1) | 4.0896(2) | 4.0902(3) | 4.0902(2) | 4.0896(1) | 4.0903(2) |
| $c$ (Å) | 14.0043(6) | 14.0060(6) | 14.0280(8) | 14.0761(13) | 14.1013(18) | 14.1227(10) | 14.0080(6) | 14.0101(11) |
| $c/a$ | 3.4261 | 3.4273 | 3.4303 | 3.4419 | 3.4476 | 3.4528 | 3.4252 | 3.4252 |
| $La_2O_2Bi$ (mol%) | 100 | 100 | 100 | 100 | 51.4 | 56.6 | 55.2 | 17.1 |
| $La_2O_3$ (mol%) | 0 | 0 | 0 | 0 | 0 | 8.3 | 18.8 | 34.8 |
| Bi (mol%) | 0 | 0 | 0 | 0 | 48.6 | 35.2 | 26.1 | 48.2 |
| $R_{wp}$ | 2.147 | 2.304 | 2.275 | 2.267 | 1.749 | 2.267 | 2.840 | 2.454 |
| $R_e$ | 1.872 | 1.869 | 1.863 | 1.751 | 1.584 | 1.780 | 1.868 | 1.892 |
| $S$ | 1.1471 | 1.2326 | 1.2212 | 1.2949 | 1.1039 | 1.2739 | 1.5205 | 1.2969 |

[a]mol%: molar fraction of the phase, $R_{wp}$: $R$-factor, $R_e$: expected $R$-factor, $S$: goodness-of-fit indicator.

## Result and discussions

Fig. 1a–f show XRD patterns for a selected series of $La_2O_2Bi$ co-sintered with each nominal composition $x_{Ca}$, $x_{Sr}$, and $x_{Ba}$ = 0.10, 0.20, whereas those with $x_{Sr}$ = 0.03, 0.16 are shown in Fig. S1†. For $x_{Ca}$ = 0.10, 0.20, and $x_{Sr}$ = 0.03, 0.10, pure $La_2O_2Bi$ phase was obtained, while impurity phases of $La_2O_3$ and Bi were observed for $x_{Sr}$ = 0.16, 0.20 and $x_{Ba}$ = 0.10, 0.20. These phase fractions as a function of $x_{Ca}$, $x_{Sr}$, and $x_{Ba}$ are summarized in Fig. 1g–i. Fig. 1j and k show $a$- and $c$- axis lengths as a function of $x_{Ca}$, $x_{Sr}$, and $x_{Ba}$, respectively. The $a$-axis length was almost constant irrespective of $x_{Ca}$, $x_{Sr}$, and $x_{Ba}$ values, while the $c$-axis length was expanded with increasing $x_{Ca}$, $x_{Sr}$, and $x_{Ba}$. The $c$-axis length was almost the same for each $x_{Ca}$ and $x_{Ba}$, and those for each $x_{Sr}$ showed a larger increase in proportion to $x_{Sr}$. From SEM-EDX measurements, Ca was fully precipitated from $La_2O_2Bi$ matrix (Fig. S2†), indicating that CaO served as an oxygen intercalant without Ca substitution with La, as was reported in other $RE_2O_2Bi$.[13,14] On the other hand, Sr was not significantly precipitated, but incorporated in $La_2O_2Bi$ matrix (Fig. S3†), representing homogeneous Sr substitution with La. Therefore, the expansion

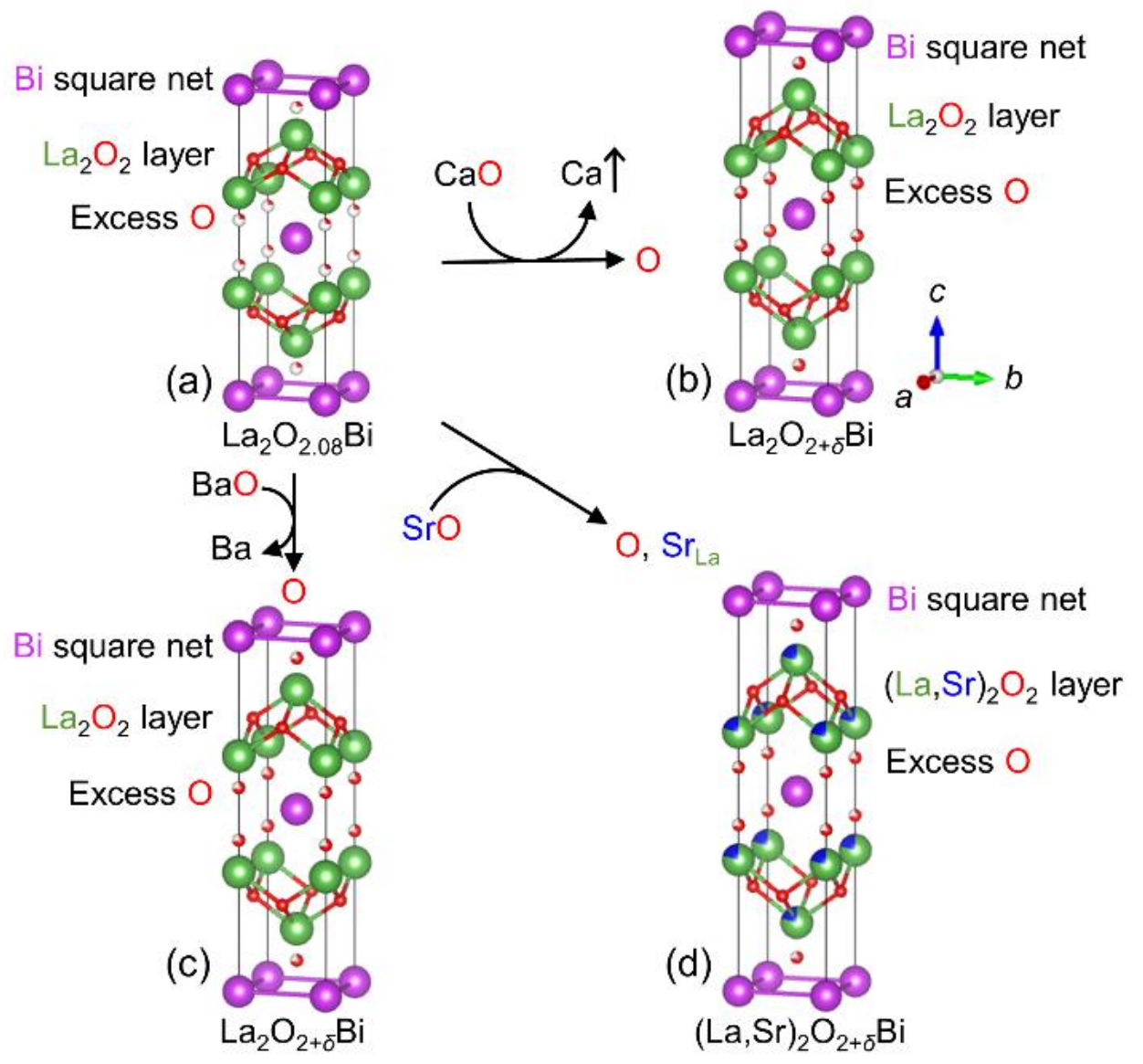

**Fig. 2** Chemical composition changes of $La_2O_2Bi$ by co-sintering with *AEO* (*AE* = Ca, Sr, and Ba). (a) Pristine $La_2O_2Bi$ before sintering with *AEO*. (b) $La_2O_2Bi$ with excess oxygen after sintering with CaO. (c) $La_2O_2Bi$ with excess oxygen accompanied by $La_2O_3$ and Ba:Bi alloy after sintering with BaO. (d) Sr substituted $La_2O_2Bi$ with excess oxygen after sintering with SrO. The site occupancy of excess oxygen in (b)–(d) was larger than that in (a).

in *c*-axis length in Fig. 1k was explained by oxygen intercalation for $La_2O_2Bi$ co-sintered with CaO and BaO, and by both oxygen intercalation and Sr substitution with La for $La_2O_2Bi$ co-sintered with SrO. Crystal structural parameters of all the samples obtained by Rietveld analyses are summarized in Table 1.

The role of *AE*O in the synthesis of $La_2O_2Bi$ is summarized in Fig. 2. All *AE*O serves as an oxidant to intercalate oxygen into $La_2O_2Bi$ resulting in the longer *c*-axis length (> 14.00 Å) than those of $La_2O_{2+\delta}Bi$ in previous study (Fig. 2a),[15] in which the excess oxygen was intercalated by controlling the nominal oxygen composition together with significant amounts of impurity phases (Fig. S4† and Table S1†). In case of CaO (Fig. 2b), the phase purity of $La_2O_2Bi$ was high probably due to Ca evaporation. In case of BaO (Fig. 2c), a large amount of Bi and $La_2O_3$ impurity phase was observed possibly due to higher reactivity of BaO than those of CaO and SrO, resulting in the formation of Ba:Bi alloy (Table S2†).[18] In case of SrO (Fig. 2d), SrO served also as a hole dopant as described below, although Bi and/or $La_2O_3$ impurity phase appeared for high $x_{Sr}$.

Here, we describe transport properties of $La_2O_2Bi$ with $x_{Ca}$ = 0.10, 0.20, and $x_{Sr}$ = 0.03, 0.10, which are almost the single phase. $La_2O_2Bi$ with $x_{Ca}$ = 0.10, 0.20, and $x_{Sr}$ = 0.03, 0.10 showed a lower resistivity at 2–300 K than those of $La_2O_{2.08}Bi$ and $La_2O_{2.20}Bi$ reported previously (Fig. 3a),[15] in addition to absence of resistivity upturn with decreasing temperature. This result suggests a larger amount of intercalated oxygen in $La_2O_2Bi$ with $x_{Ca}$ = 0.10, 0.20 and $x_{Sr}$ = 0.03, 0.10 than that in $La_2O_{2.20}Bi$, as was exemplified in the longer *c*-axis length (Fig. 1k). $La_2O_2Bi$ with $x_{Sr}$ = 0.10 showed the lowest resistivity, indicating hole carrier doping by Sr substitution, in spite of absence of superconducting transition down to 0.07 K (Fig. 3a inset) probably due to significantly lower carrier density of $La_2O_2Bi$ in comparison with other superconducting $RE_2O_2Bi$ like $Y_2O_2Bi$ (1.0

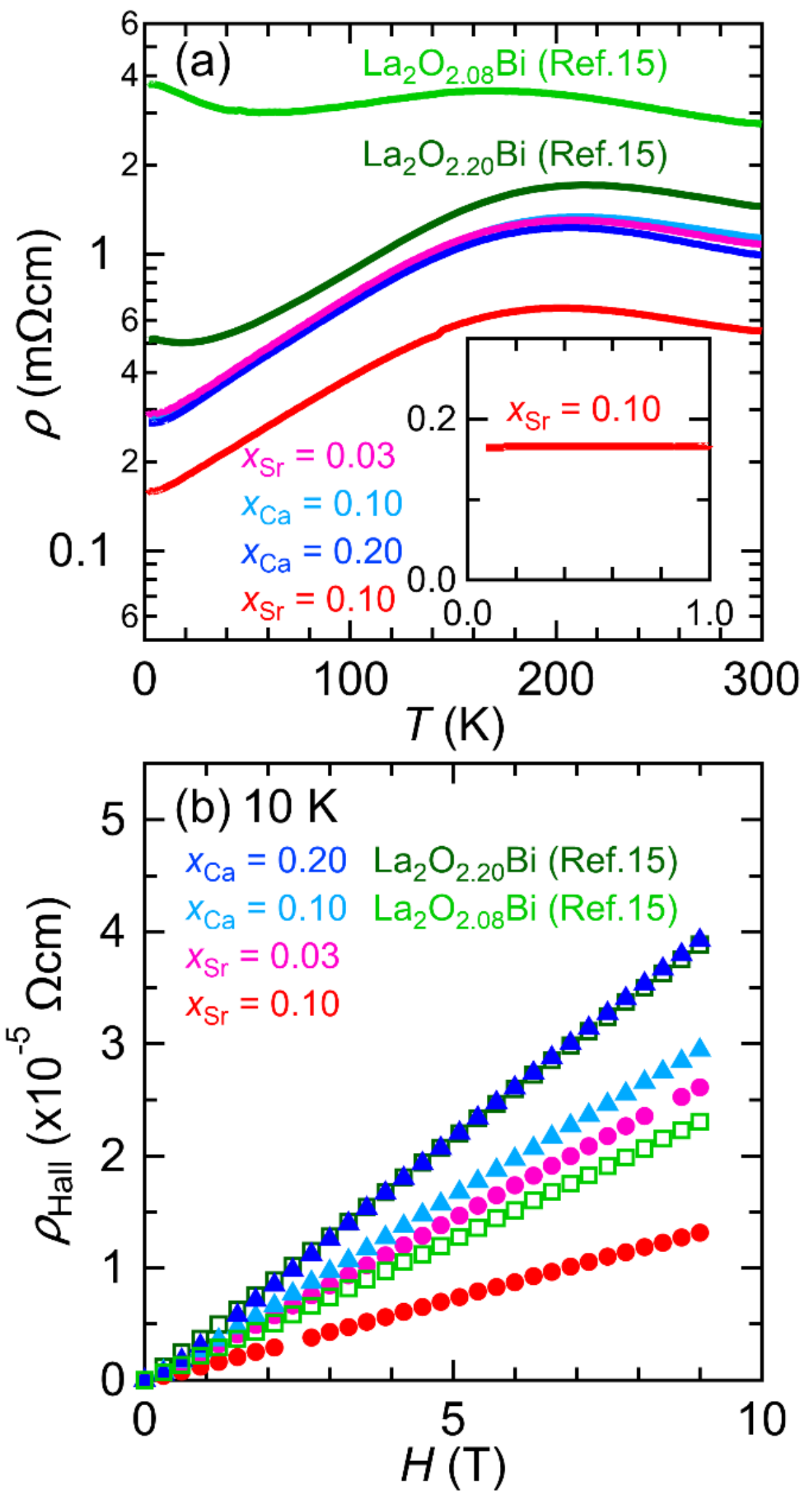

**Fig. 3** Temperature dependence of (a) resistivity and (b) magnetic field dependence of Hall resistivity for $La_2O_{2.08}Bi$, $La_2O_{2.20}Bi$, $La_2O_2Bi$ with $x_{Ca}$ = 0.10, 0.20, and $La_2O_2Bi$ with $x_{Sr}$ = 0.03, 0.10.[15] The inset in (a) shows the resistivity for $La_2O_2Bi$ with $x_{Sr}$ = 0.10 below 1.0 K.

× $10^{23}$ $cm^{-3}$ at 100 K).[12] Magnetic field dependence of Hall resistivity for $La_2O_2Bi$ with $x_{Ca}$ = 0.10, 0.20 and $x_{Sr}$ = 0.03, 0.10 at 10 K is shown in Fig. 3b. The Hall resistivity was positively proportional to magnetic field, corresponding to the hole carrier conduction. The similar slope of the Hall resistivity for $La_2O_2Bi$ with $x_{Ca}$ = 0.10, 0.20, and $x_{Sr}$ = 0.03 to those of $La_2O_{2.08}Bi$ and $La_2O_{2.20}Bi$ indicated their similar amounts of hole carrier density, representing that the excess oxygen was not effective carrier dopant, as was discussed in the previous study.[15] On the other hand, the smaller slope for $La_2O_2Bi$ with $x_{Sr}$ = 0.10 was caused by the increased hole carrier density due to Sr substitution with La, as was observed in other layered compounds such as $La_{2-x}Sr_xCuO_4$ and $La_{1-x}Sr_xNiAsO$.[19,20]

The resistivity and mobility at 10 K are summarized as a function of hole carrier density in Fig. 4a and b, respectively. The resistivity was decreased by co-sintering with CaO or SrO. For the case of CaO, the decrease in resistivity was mainly caused by the enhanced mobility up to 150 $cm^2V^{-1}s^{-1}$ for $La_2O_2Bi$ with $x_{Ca}$ = 0.20, while the hole carrier density were similar to those of

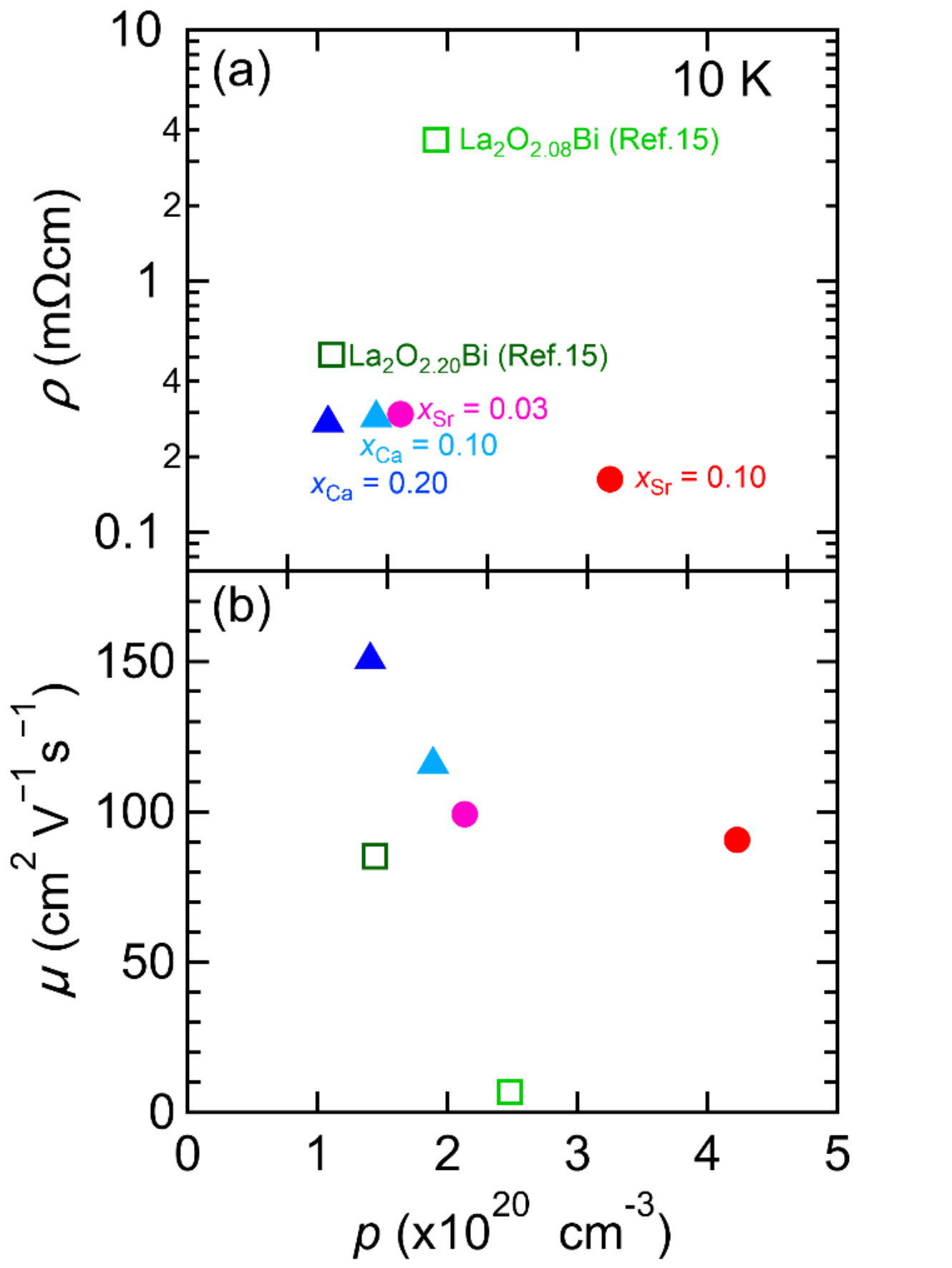

**Fig. 4** (a) Resistivity and (b) hole mobility at 10 K as a function of hole carrier density for $La_2O_{2.08}Bi$, $La_2O_{2.20}Bi$, $La_2O_2Bi$ with $x_{Ca}$ = 0.10, 0.20, and $La_2O_2Bi$ with $x_{Sr}$ = 0.03, 0.10.[15]

$La_2O_{2.08}Bi$ and $La_2O_{2.20}Bi$ (Fig. 4b).[15] This result suggests that excess oxygen served to increase the hole carrier mobility possibly due to the enhanced two-dimensionality of Bi square net.[21] For the case of SrO, on the other hand, the lowest resistivity was obtained for $La_2O_2Bi$ with $x_{Sr}$ = 0.10 by heavy hole carrier doping (4.2 × $10^{20}$ $cm^{-3}$) while the hole carrier mobility was similar to those of $La_2O_{2.08}Bi$ and $La_2O_{2.20}Bi$ (Fig. 4b).[15] Accordingly, there are two approaches to decrease resistivity of $La_2O_2Bi$ through increasing either carrier mobility or hole carrier density.

## Conclusions

In this study, $La_2O_2Bi$ with longer *c*-axis length than previous study (Ref.15) was obtained by co-sintering with CaO, SrO, and BaO, which served as the oxygen intercalant. The use of CaO yielded pure phase of $La_2O_2Bi$ with the enhanced hole mobility up to 150 $cm^2V^{-1}s^{-1}$ at 10 K. SrO served as not only the oxygen intercalant but also as the hole dopant by substituting La, realizing the higher electric conductivity of $La_2O_2Bi$ than ever.

## Acknowledgements

The authors acknowledge Dr. D. Oka for technical support. This study was supported by JSPS KAKENHI (No. 26105002) and Yazaki Memorial Foundation for Science and Technology.